\documentclass{article}

\usepackage{arxiv}

\usepackage[utf8]{inputenc} 
\usepackage[T1]{fontenc}    
\usepackage{hyperref}       
\usepackage{url}            
\usepackage{booktabs}       
\usepackage{amsfonts}       
\usepackage{nicefrac}       
\usepackage{microtype}      
\usepackage{lipsum}
\usepackage{graphicx}
\usepackage{xcolor}
\graphicspath{ {./images/} }
\usepackage{caption}
\usepackage[linesnumbered]{algorithm2e}
\usepackage{amsmath}
\usepackage[stable]{footmisc}
\usepackage{fixltx2e}

\title{Representation Learning of Image Schema}

\author{
 Fajrian Yunus \\
  ISIR\\
  Sorbonne University\\
  Paris, France \\
  \texttt{fajrian.yunus@gmail.com} \\
   \And
 Chlo\'{e} Clavel \\
  LTCI\\
  T\'{e}l\'{e}com-Paris, IP-Paris\\
  Paris, France \\
  \texttt{chloe.clavel@telecom-paristech.fr} \\
  \And
 Catherine Pelachaud \\
  ISIR\\
  CNRS / Sorbonne University\\
  Paris, France \\
  \texttt{catherine.pelachaud@sorbonne-universite.fr} \\
}

\begin{document}
\maketitle

\begin{abstract}

Image schema is a recurrent pattern of reasoning where one entity is mapped into another. Image schema is similar to conceptual metaphor and is also related to metaphoric gesture. Our main goal is to generate metaphoric gestures for an Embodied Conversational Agent. 

We propose a technique to learn the vector representation of image schemas. As far as we are aware of, this is the first work which addresses that problem. Our technique uses Ravenet et al\textquotesingle s algorithm which we use to compute the image schemas from the text input and  also BERT and SenseBERT which we use as the base word embedding technique to calculate the final vector representation of the image schema. Our representation learning technique works by clustering: word embedding vectors which belong to the same image schema should be relatively closer to each other, and thus form a cluster.

With the image schemas representable as vectors, it also becomes possible to have a notion that some image schemas are closer or more similar to each other than to the others because the distance between the vectors is a proxy of the dissimilarity between the corresponding image schemas. Therefore, after obtaining the vector representation of the image schemas, we calculate the distances between those vectors. Based on these, we create visualizations to illustrate the relative distances between the different image schemas.

\end{abstract}

\section{Introduction}

Image schema is a recurrent pattern of reasoning where one entity is mapped into another~\cite{johnson2013body}. Image schema itself is similar to conceptual metaphor~\cite{lakoff1980conceptual} where human talks about one thing by using another object which has similar properties. Cienki~\cite{cienki2013image} suggests that this mapping mechanism is how human produces metaphoric gestures. Based on this relationship between image schema and metaphoric gestures, Ravenet et al~\cite{ravenet2018automatic,ravenet2018automating} develop a technique to compute metaphoric gestures from a free-form text input through the image schema. The technique works in two steps: computing the image schemas from the free-form text input and computing the gestures from the image schemas.

In a separate but related development, there have been works that use neural network to generate communicative gestures for Embodied Conversational Agents ECAs. Some of these machine-learning based works generate the gestures according to a free-form text input where the text input is processed by using a word embedding technique (e.g. \cite{kucherenko2020gesticulator} and \cite{ahuja2019language2pose}). It is an interesting possibility to use image schema in such neural network systems. However neural network needs vectors as its input. Therefore, it is necessary to find a vector representation of the image schemas.

Machine learning community has undertaken various works to represent data which is in an arbitrary format as vectors so that such data can be used in machine learning. This line of works is called ``representation learning''. Word embedding techniques are representation learning techniques to represent a word as a vector. These word embedding techniques have a property that two similar words are mapped into two nearby vectors, although different word embedding techniques have different notions on what makes two words similar. Incidentally, with word embedding vectors, it is also possible to quantify the similarity between two words because the distance between the vectors is a proxy of the similarity between the words.

In this work, we propose some improvements to Ravenet et al\textquotesingle s algorithm~\cite{ravenet2018automatic,ravenet2018automating} on the computation of image schema from a free-form text input. After that, going to the main focus of this work, we propose a method to compute the embedding vectors of the image schemas. As far as we are aware of, this is the first work which addressed the problem of representation learning for image schemas. Once the image schemas are representable as vectors, it also becomes possible to measure their distances from each other. Therefore, we also propose a method to calculate the distance between different image schemas. Additionally, we show visualizations of the relative distances between different image schemas so that we can see which image schemas are relatively close to each other and which image schemas are relatively far from the others.

We organize this paper as following. In Section \ref{sec:background-image_schema}, we explain the relevant background knowledge, namely image schema and its relationship with metaphoric gestures (Section \ref{sec:background-image_schema}) and WordNet (Section \ref{sec:wordnet}). In Section \ref{sec:existing_image_schema_competition}, we explain the prior works which deal with both image schema and metaphoric gesture, mainly the work of Ravenet et al~\cite{ravenet2018automatic,ravenet2018automating}. In Section \ref{sec:word-embeddings}, we explain word embedding techniques, including the techniques which will become the basis of our work: BERT~\cite{devlin2018bert} and SenseBERT~\cite{levine2020sensebert}. In Section \ref{sec:limitation_of_ravenet_algo}, we explain the limitations of Ravenet et al\textquotesingle s algorithm and our proposed enhancements. In Section \ref{sec:proposed_method}, we explain our proposed methods to compute the image schema embedding and to compute the distance between different image schemas. In Section \ref{sec:experiment}, we put our proposed methods into experiments. In Section \ref{sec:discussion}, we discuss the interpretation of the results from our experiments. Finally, in Sections \ref{sec:conclusion} and \ref{sec:future_work}, we close this paper with our conclusion and the future work.

\section{Background} \label{sec:background}

\subsection{Image Schema and Metaphoric Gestures} \label{sec:background-image_schema}
Johnson~\cite{johnson2013body} defines image schema as a recurrent pattern of reasoning where one entity is mapped into another. For example, ``culture'' can be mapped into ``container'' by thinking that some people belong to the same culture while some other people do not. Therefore, culture shares the same property with a container that both of them have a boundary.

Image schema is similar to conceptual metaphor from Lakoff and Johnson~\cite{lakoff1980conceptual} where human talks about one thing by using another object which has similar properties. For example, in a metaphor ``love is a journey'', ``love'' is imagined to consist of the starting point, the destination, and the path which links both the starting point and the destination. This phenomenon is also observed in our language. In English, we can say ``big idea'' to mean an idea which has the potential to make a significant impact. However, ``big'' itself is a property normally used for a concrete object. ``Idea'' is an abstract object, and therefore it is neither big nor small. As such, ``idea'' has to be mentally mapped into a concrete object which has a physical size.

This ``metaphorizing'' is relevant because the conceptualization hypothesis states that the way human represents the world in their mind is constrained by the human\textquotesingle s physical body~\cite{wilson2013embodied}, which means there is a need to map an abstract entity into a concrete entity. Metaphor can even affect the physical body movement unconsciously. Miles et al~\cite{miles2010moving} find in their experiment that their participants lean forward while thinking about future events. On the other hand, they lean backward while thinking about past events.

\begin{figure}
  \centering
  \includegraphics[scale=1]{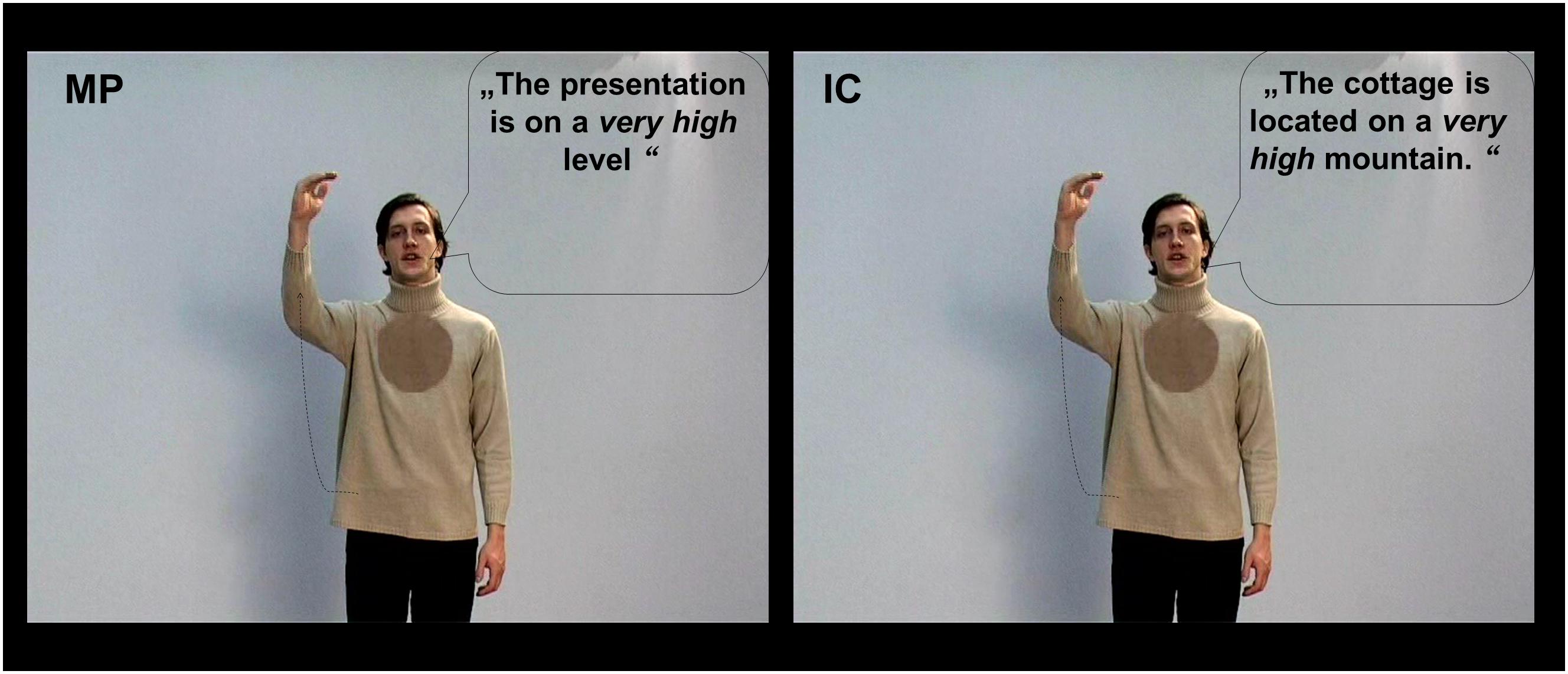}
  \caption{He et al~\cite{he2018role}\textquotesingle example of metaphoric gesture (the left part, labelled with ``MP''). It can be seen here that the sophistication of the presentation is thought as a concrete object which has the height property (e.g. a mountain), as shown in the right part (labelled with ``IC'')}
  \label{fig:he2018_metaphoric_gesture_example}
\end{figure}

Cienki~\cite{cienki2013image} suggests that this mapping mechanism is how human produces metaphoric gestures. Metaphoric gesture itself is a gesture which depicts an abstract concept through the aforementioned ``metaphorizing'' process. He et al~\cite{he2018role} show an example where a person rises his hand to describe the high level (of the quality/sophistication) of a presentation (see Figure \ref{fig:he2018_metaphoric_gesture_example}). In this case, through ``metaphorizing'', the quality/sophistication of the presentation is thought as something which has the height property (e.g. a mountain).

L\"{u}cking et al~\cite{lucking2016finding} and Cienki~\cite{cienki2008image} experimentally investigate the presence of the relationship between image schema and metaphoric gestures. L\"{u}cking et al~\cite{lucking2016finding} do an experiment where they ask the participants to perform gestures to manifest various terms, including image-schema terms, by using hand and arm movements. L\"{u}cking et al find that the participants tend to perform similar gestures for some stimulus terms. In a separate work, Cienki~\cite{cienki2008image} experimentally investigates the concordance of image schema in different modalities. The idea of Cienki\textquotesingle s experiment is to have the same ``message'' but in different combination of modalities (visual, audio, or transcript), and then the participants are asked to label those ``messages'' with the image schema. The ``message'' comes from a video which is then processed to remove certain modalities (e.g. transcribing the speech and retaining the audio but removing the visuals). It is observed in Cienki\textquotesingle s experiments that certain gesture types show a greater concordance between the different combination of modalities. The findings from the two aforementioned works suggest that there is indeed a relationship between image schema and metaphoric gestures.

\subsection{WordNet} \label{sec:wordnet}
\begin{figure}
    \centering
	\includegraphics[scale=0.9]{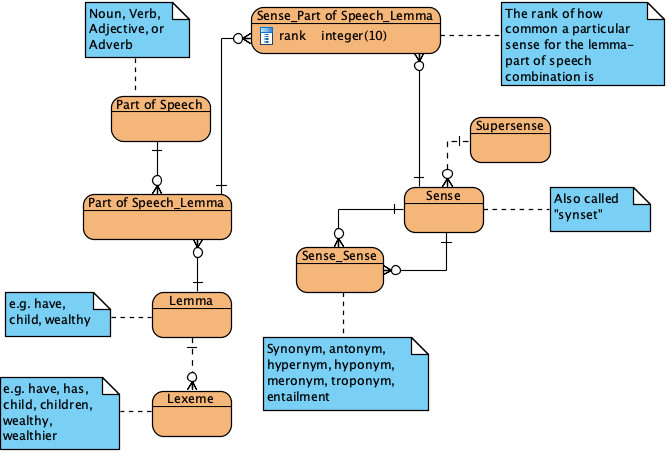}
	\caption{WordNet\textquotesingle s schema in Entity-Relationship diagram}
	\label{fig:wordnet_er_diagram}
\end{figure}

WordNet~\cite{miller1995wordnet} is a lexical database of English. It is organized as a directed graph. Each node is a ``sense'' (i.e. meaning), which is called ``synset''. Each sense has uniquely one part-of-speech tag (i.e. noun, verb, adjective, or adverb), but can have multiple lemmas associated with the sense. For example, a verb with the sense of being cognizant or aware of a fact or a specific piece of information can be represented by one of these lemmas: ``know'', ``cognize'', ``cognise''. WordNet only keeps the lemma; therefore, the conjugation differences are eliminated. For example, ``have'', ``has'', and ``had'' belong to the same lemma. ``Child'' and ``children'' also belong to the same lemma.  Similarly, ``wealthy'', ``wealthier'', and ``wealthiest'' also belong to the same lemma. However, different spellings of the same word are considered to be different lemmas. For example, ``organize'' and ``organise'' are different lemmas. A combination of a lemma and a part-of-speech-tag can have multiple possible senses. For example, the noun ``fan'' has three possible senses: a device for creating a current of air by movement of a surface/surfaces, an enthusiastic devotee of sports, or an ardent follower and admirer. The senses are ordered from the most common sense to the least common sense. Two senses might have an edge connecting them. There are several edge types: ``synonym'', ``antonym'', ``hypernym'', ``hyponym'', ``meronym'' (one is a part of another), ``troponym'' (i.e. manner of doing), and ``entailment''.  The WordNet\textquotesingle s schema can be seen in Figure \ref{fig:wordnet_er_diagram}. Each WordNet sense belongs to uniquely one WordNet supersense, but one WordNet supersense can have many senses. The list of the supersenses is furnished in Table \ref{tab:wordnet_supersenses}. It should be noted that supersense is a different property from sense. Sense is not related to supersense via the WordNet edges. We also furnish the statistics of the senses and the lemmas in WordNet in Tables \ref{tab:wordnet_sense_count} and \ref{tab:wordnet_lemma_count}. It can be seen in the Table \ref{tab:wordnet_sense_count} that noun senses far outnumber all other senses. Interestingly, there are also far more adjective senses than adverb senses, even though adjective and adverb are strongly related in English. Similarly, in Table \ref{tab:wordnet_lemma_count}, we observe again that the nouns far outnumber the rest. Similarly, the adjectives also outnumber the adverbs in terms of lemma count.

\begin{table}
	\begin{center}
		\captionsetup{justification=centering}
		\caption{The WordNet\textquotesingle s supersenses}
		\label{tab:wordnet_supersenses}
		\begin{tabular}{|c|c|c|c|} 
			\hline
			Adjective & Adverb & Noun & Verb \\
			\hline
			all & all & act & body \\
			pert &  & animal & change \\
			ppl &  & artifact & cognition \\
			&  & attribute & communication \\
			&  & body & competition \\
			&  & cognition & consumption \\
			&  & communication & contact \\
			&  & event & creation \\
			&  & feeling & emotion \\
			&  & food & Framestext \\
			&  & group & motion \\
			&  & location & perception \\
			&  & motive & possession \\
			&  & object & social \\
			&  & person & stative \\
			&  & phenomenon & weather \\
			&  & plant &  \\
			&  & possession &  \\
			&  & process &  \\
			&  & quantity &  \\
			&  & relation &  \\
			&  & shape &  \\
			&  & state &  \\
			&  & substance &  \\
			&  & time &  \\
			&  & Tops &  \\
			\hline
		\end{tabular}
	\end{center}
\end{table}

\begin{table}
	\begin{center}
		\captionsetup{justification=centering}
		\caption{WordNet 3.0\textquotesingle s sense count}
		\label{tab:wordnet_sense_count}
		\begin{tabular}{|c|c|} 
			\hline
			Type & Count \\
			\hline
			Noun & 82115 \\
			Verb & 13767 \\
			Adjective & 18156 \\
			Adverb & 3621 \\
			\hline
		\end{tabular}
	\end{center}
\end{table}

\begin{table}
	\begin{center}
		\captionsetup{justification=centering}
		\caption{WordNet 3.0\textquotesingle s lemma count. A lemma may cover several part-of-speech types}
		\label{tab:wordnet_lemma_count}
		\begin{tabular}{|c|c|} 
			\hline
			Type & Count \\
			\hline
			Noun & 117798 \\
			Verb & 11529 \\
			Adjective & 21479 \\
			Adverb & 4481 \\
			All & 147306 \\
			\hline
		\end{tabular}
	\end{center}
\end{table}

\section{Related Work} \label{sec:related_work}

\subsection{Existing Image Schema Computation} \label{sec:existing_image_schema_competition}

L\"{u}cking et al~\cite{lucking2016finding} and Cienki~\cite{cienki2008image} experimentally investigate the presence of the relationship between image schema and metaphoric gestures (see Section \ref{sec:background-image_schema}). However, in those two works, the image schema is already known beforehand. Therefore there remains a question on how to actually compute the image schema from a free-form text. Ravenet et al~\cite{ravenet2018automatic,ravenet2018automating} develop a technique to compute metaphoric gestures from a free-form text input through the image schema. The algorithm works in two steps: first by computing the image schema from the free-form text input, and secondly by computing the gestures according to the resulting image schema. Thus, the first part of Ravenet et al\textquotesingle s algorithm solves the problem on how to compute the image schema from a free-form text input.

The technique of Ravenet et al to compute the image schema from a free-form text input works by first parsing the text to get the corresponding parts of speech. This parsing is done by using the Stanford POS Tagger~\cite{toutanova2003feature}. Then, the technique gets the lemma of the word. For example, ``walk'', ``walks'', and ``walked'' will be converted into ``walk''. Similarly, both ``hand'' and ``hands'' will be converted into ``hand''. It should be noted, however, gendered nouns exist but are rare in English (e.g. ``actor''-``actress'', ``governor''-``governess''). These gendered nouns are considered as different words. Then, based on the lemmas of the word (e.g. ``walk'' for ``walks'') and the part of speech (e.g. verb), the technique decides the correct sense by using simplified Lesk algorithm~\cite{lesk1986automatic}. The simplified Lesk algorithm works by calculating the size of the intersection between the possible lemmas of each sense and the lemmas in the rest of the sentence. The sense whose size of the intersection is the largest is considered to be the correct sense. Once the correct sense is decided, then the WordNet sense graph is traversed through the hypernym edges until it finds one of the ``terminal senses/nodes''. The traversal is done by a depth first search traversal. The algorithm has a mapping set by static rules which map a set of senses/nodes to an image schema. Thus, the relationship between the image schema and the terminal senses/nodes is one-to-many. The image schema is decided based on this terminal WordNet sense. If there are multiple words with image schemas in a phrase, then we prioritize the word which is tagged to have pitch accent. The pitch accent is relevant because the stroke often happens at around the pitch accent~\cite{kendon2004gesture}, and thus the gesture is more likely to depict the word where the pitch accent is. The algorithm is furnished in Algorithm \ref{algo:existing_meaning_miner_algo}.

\begin{algorithm}
	\DontPrintSemicolon 
	$split\ the\ sentence\ into\ phrases\ and\ the\ POS\ tags$;
	
	\For{phrase in all\ phrases} { \label{algo_line:per_phrase_loop}
		
		\For{word\ and\ POS\ tag\ in\ the\ phrase} {
			
			\If {has\ pitch\ accent\ or\ is\ adjective\ or\ is\ adverb} {
				
				$mark\ this\ word\ as\ priority$
				
			}
			
			$lemmas \gets getPossibleLemmas(word, POS\_tag)$;
			
			\For{lemma\ in\ lemmas} { \label{algo_line:lemma_loop}
				$mostLikelySense \gets simplifiedLeskAlgorithm(lemma, POS\_tag)$; \label{algo_line:lesk}
				
				$//traverse\ the\ WordNet\ graph\ via\ hypernym\ edges$;
				$imageSchemas += getImageSchemaByTraversingWordNetGraph(mostLikelySense)$; \label{algo_line:hypernym_traversal}
			}
			
			\If {has\ image\ schema} {
				$chosenImageSchema = imageSchemas[0]$
			}
		}
		
		\For{word\ and\ POS\ tag\ in\ the\ phrase} {
			
			\uIf {priority\ word\ has\ image\ schema} {
				$choose\ this\ image\ schema$;
			}
			\uElseIf {non\ priority\ word\ has\ image\ schema} {
				$choose\ this\ image\ schema$;
			}
			\uElse {
				$this\ phrase\ has\ no\ image\ schema$;
			}
			
		}    
		
	}
	
	\caption{Ravenet et al\textquotesingle s algorithm to compute image schema~\cite{ravenet2018automatic,ravenet2018automating} from a free-form text input}
	\label{algo:existing_meaning_miner_algo}
\end{algorithm}

\subsection{Word Embedding} \label{sec:word-embeddings}
One recent trend in natural language processing is the use of word embedding to represent a word in the form of a vector. This vector representation can then be used in various machine learning models for various problems. For example, when BERT~\cite{devlin2018bert} is being proposed, it is tested on natural language understanding tasks, questions answering tasks, and sentence continuation tasks.

Word embedding has a property that two words which have similar meanings have their vector representations also close to each other. It should be noted, however, word embedding is not a single technique, instead it is better seen as a class of techniques which shares the same principle, but is implemented differently. Especially, different techniques have different notions of similarity and how this similarity is translated into distance in the vector space. The distance between the embedding vectors are calculated by using the euclidean distance or the cosine dissimilarity.

Word2Vec~\cite{mikolov2013distributed} is the pioneer of word embedding. It uses the notion that two words are similar if they are surrounded by the same words. There are two variants of Word2Vec. The first one is basically a neural network which does a ``fill in the blank'' task. Given an n-gram whose middle word is hidden, the network is to guess the hidden word. This is called ``Continuous Bag Of Words'' (CBOW) approach. The second variant is the reverse: given the middle word, the network is to guess the rest of the n-gram. This is called ``Skip-Gram'' approach. The result of any of these approaches is that, two words which are surrounded by similar words (in an n-gram) will have the corresponding vectors also close to each other.

GloVe~\cite{pennington2014glove} follows a similar notion. However, While Word2Vec treats different n-grams as different samples, GloVe learns from the global data of co-occurrences. GloVe works by doing an optimization such that the distance between the two corresponding vectors are minimized when their probabilities of co-occurrence with other words are similar. Effectively, it means two words are similar if they tend to co-occur with similar words.

\begin{figure}
    \begin{minipage}{\textwidth}
        \centering
    	\includegraphics[scale=0.4]{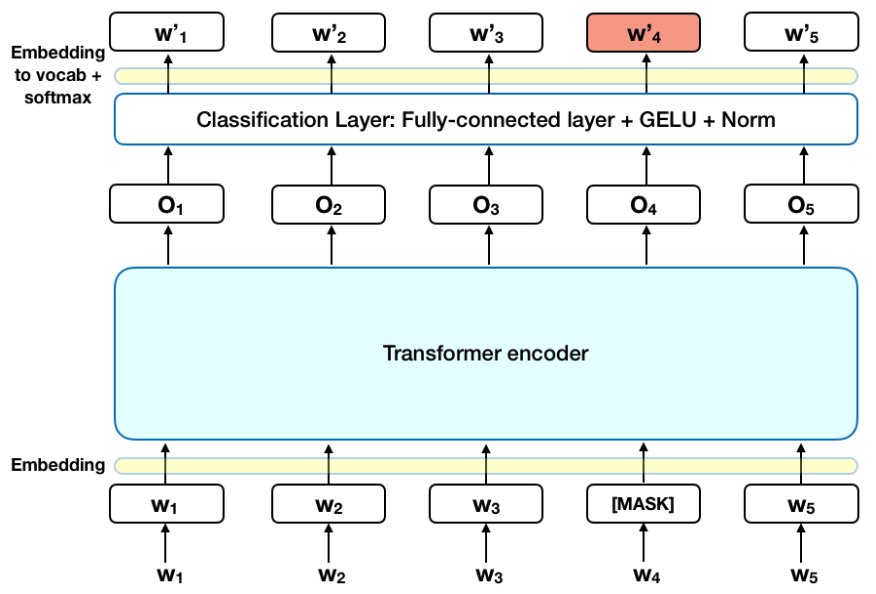}
    	\caption{The ``fill in the blank'' training of BERT~\cite{devlin2018bert}~\footnote{The schema is from \url{https://towardsdatascience.com/bert-explained-state-of-the-art-language-model-for-nlp-f8b21a9b6270}}. The network learns to predict $w_4$.}
    	\label{fig:bert_training_fill_in_the_blank}
	\end{minipage}
\end{figure}

Both Word2Vec and Glove have ``static'' embedding. A word always has the same embedding, no matter the context where it appears. However, one word can have different meanings depending on the context. For example, ``fan'' can mean a follower or admirer (e.g. ``she is a fan of Justin Bieber'') or an instrument to displace air (e.g. ``I need a fan because my room is hot''). BERT~\cite{devlin2018bert} on the other hand, yields different embeddings for the same word depending on the context. This is called ``contextual word embedding''. BERT training works in two ways. The first one is that it takes the sentence as the input, with a few words being hidden, and then the network learns to re-produce the same sentence including the hidden words. Here, the network learns to ``fill in the blank'' (see Figure \ref{fig:bert_training_fill_in_the_blank}). The second one is that the network is given two sentences (some words are hidden), and the network learns to re-produce the complete sentences and to indicate whether the two sentences are consecutive. Here, the network learns if the context of the two sentences are related. Thus, it can be seen that unlike Word2Vec or GloVe which learn the word embedding in isolation, BERT learns the context together. Consequently, the same word will have different corresponding vectors depending on the sentence.

\begin{figure}
    \centering
	\includegraphics[scale=0.4]{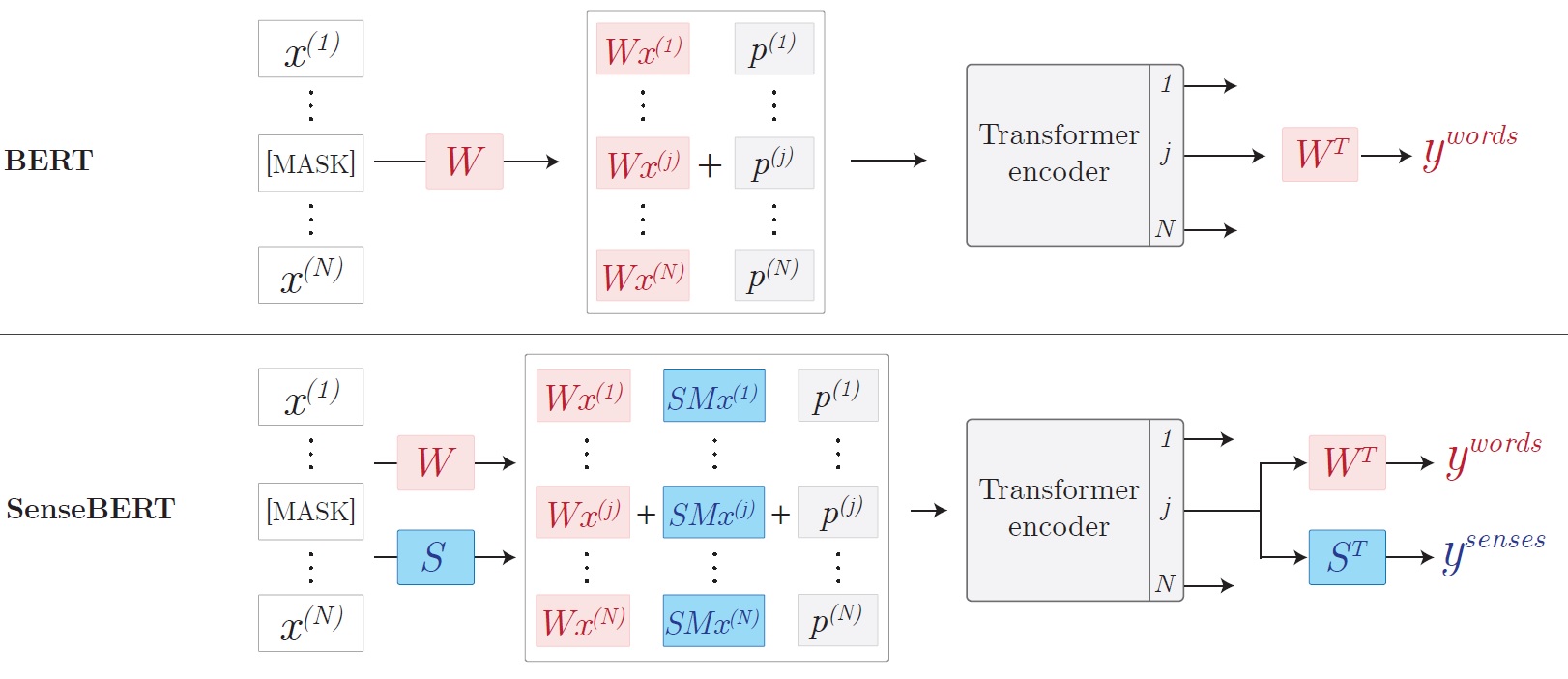}
	\caption{The difference between BERT and SenseBERT~\cite{levine2020sensebert}. Unlike BERT, during the ``fill in the blank'' of SenseBERT, it tries to predict the WordNet supersense as well (see Table \ref{tab:wordnet_supersenses}).}
	\label{fig:bert_vs_sensebert_schema}
\end{figure}

SenseBERT~\cite{levine2020sensebert} is an extension BERT by also taking into account the similarities in WordNet in the training. Therefore, unlike BERT, SenseBERT explicitly works at the word sense level. Specifically, unlike BERT which learns the words only, SenseBERT learns both the words and the corresponding supersenses. Therefore, it can be said that SenseBERT consider two words to be more similar when they have the same corresponding supersense. Levine et al show that SenseBERT outperforms BERT on both SemEval word sense disambiguation tasks and ``word in context'' tasks. The schema depicting the difference between BERT and SenseBERT is available in Figure \ref{fig:bert_vs_sensebert_schema}.

We can see from these works that word embedding can be used to represent text and there are many variants of word embedding. Therefore, we have three research questions we aim to address. The first one is how can we use word embedding to represent an image schema as a vector. The second one is how do we compare the word embedding we choose against the alternatives. However, if we have the image schema vectors, then the distances between different image schemas also become computable, and thus it makes sense to ask if certain image schemas are closer to each other. Therefore, in the third research question, we want to investigate which image schemas are close to each other.

\section{The Limitations of the Ravenet et al\textquotesingle s Algorithm and Our Proposed Enhancement} \label{sec:limitation_of_ravenet_algo}

In this section, we discuss the limitations of Ravenet et al\textquotesingle s algorithm on the image schema computation and our proposal on how to address them. The first one is the limitation of the Lesk algorithm which Ravenet et al use in their algorithm (see Algorithm \ref{algo:existing_meaning_miner_algo} Line \ref{algo_line:lesk}) to disambiguate the word. In Section \ref{sec:lesk_algorithm_wsd}, we explain that some researchers compare the Lesk algorithm against other word sense disambiguation techniques and find that a far simpler algorithm, namely always choosing the WordNet\textquotesingle s first sense, outperforms Lesk algorithm in several different experiments. The second one is the limitation of the hypernym-only WordNet graph traversal (see Algorithm \ref{algo:existing_meaning_miner_algo} Line \ref{algo_line:hypernym_traversal}). In Section \ref{sec:hypernym_only_wordnet_graph_traversal}, we explain why using only the hypernym edges is not sufficient and how we propose to address this issue.

\subsection{Lesk Algorithm For Word Sense Disambiguation} \label{sec:lesk_algorithm_wsd}
Ravenet et al~\cite{ravenet2018automatic,ravenet2018automating} use Lesk algorithm for word sense disambiguation in their image schema computation technique. However, Lesk algorithm is not the only word sense disambiguation techniques. Raganato et al~\cite{raganato2017word} compare various such techniques. The techniques include those which are based on machine learning, namely IMS~\cite{zhong2010makes}, IMS+embeddings~\cite{taghipour2015semi,rothe2015autoextend,iacobacci2016embeddings}, Context2Vec~\cite{melamud2016context2vec}, and also knowledge-based techniques, namely Lesk~\cite{lesk1986automatic}, UKB~\cite{agirre2009personalizing,agirre2014random}, and Babelfly~\cite{moro2014entity}. The machine-learning based techniques are trained on SemCor~\cite{miller1994using} and OMSTI~\cite{taghipour2015one} corpora. Raganato et al also add two simple baseline techniques, namely Most Frequent Sense (MFS) technique which always chooses the most common sense in the training corpus and WordNet First Sense technique which always chooses the first WordNet sense. It should be remembered that the first WordNet sense is the most common sense. Raganato et al compare those techniques in a uniform setting. They use the data from Senseval~\cite{edmonds2001senseval,snyder2004english} and SemEval~\cite{pradhan2007semeval,navigli2013semeval,moro2015semeval} corpora for testing. All the data have the part-of-speech tagging done by using Stanford CoreNLP toolkit~\cite{manning2014stanford}. All the corpora include WordNet sense annotations. They use F-Measure to quantify the performance of the Word Sense Disambiguation techniques. The results of Raganato et al\textquotesingle s experiments are displayed in Table \ref{tab:wsd_comparison}.

As we can see in Table \ref{tab:wsd_comparison}, the Lesk algorithm always performs poorly, even compared to the other knowledge-based techniques. Interestingly, the very simple WordNet 1st Sense always outperforms the Lesk algorithm. Ravenet et al\textquotesingle s algorithm itself relies on WordNet, which means having access to WordNet is anyway necessary. Therefore, we propose to simplify the problem by always choosing the WordNet First Sense as the most likely sense. It should be noted, however, in the past several years there have been new word sense disambiguation techniques which were not tested in the aforementioned Raganato et al\textquotesingle s experiments. We will consider the new word sense disambiguation techniques in the future work.

\begin{table}
	\begin{center}
		\captionsetup{justification=centering}
		\caption{F-score comparison of the word sense disambiguation techniques~\cite{raganato2017word}}
		\label{tab:wsd_comparison}
		\begin{tabular}{|c|c|c|c|c|c|} 
			\hline
			Corpus & Range of Machine & Range of Knowledge & Lesk & MFS & WordNet \\
			& Learning Techniques & Based Techniques & Algorithm & & 1st Sense \\
			\hline
			Senseval-2 & 70.8 - 73.3 & 50.6 - 67.0 & 50.6 & 66.5 & 66.8 \\
			Senseval-3 & 68.2 - 69.6 & 44.5 - 63.7 & 44.5 & 60.4 & 66.2 \\
			SemEval-07 & 58.5 - 61.5 & 32.0 - 56.7 & 32.0 & 52.3 & 55.2 \\
			SemEval-13 & 65.0 - 67.2 & 53.6 - 66.4 & 53.6 & 62.6 & 63.0 \\
			SemEval-15 & 64.2 - 71.7 & 51.0 - 70.3 & 51.0 & 64.2 & 67.8 \\ 
			\hline
		\end{tabular}
	\end{center}
\end{table}

\subsection{Hypernym-Only WordNet Graph Traversal} \label{sec:hypernym_only_wordnet_graph_traversal}

As can be seen in Algorithm \ref{algo:existing_meaning_miner_algo} Line \ref{algo_line:hypernym_traversal}, the WordNet graph traversal in the original Ravenet et al\textquotesingle s algorithm is only done through the hypernym edges. However, WordNet has hypernym edge for noun and verb senses only~\cite{miller1995wordnet}. Therefore, we need a new method to traverse the WordNet graph. For adjective, we use the synonym relationship instead (e.g. ``essential'' is a synonym of ``important''). It should be noted, however, synonym edge is bidirectional. $x$ is a synonym of $y$ if and only if $y$ is a synonym of $x$. For adverb, we get the corresponding adjective by using the ``derived from adjective'' edge (e.g. ``importantly'' is derived from the adjective ``important''), then we do as the aforementioned treatment of adjectives. For verb, we also use the troponym edge. Troponym itself means a manner of doing an action. For example, both ``walk'' and ``fly'' are troponyms of ``move''. It should be noted that for verb, hypernym and troponym are the inverse. $x$ is a hypernym of $y$ if and only if $y$ is a troponym of $x$. The algorithms are furnished in Algorithm \ref{algo:proposed_image_schema_algo}. We also present the statistics of about their connections in Table \ref{tab:wordnet_unordered_connected_sense_count}. It can be seen from the number of the connections and the number of senses/nodes for each part of speech (Table \ref{tab:wordnet_sense_count}) that the graph is very sparse: there are even less edges than nodes. It means, from any given node, there are only a few other nodes we can reach.

\begin{algorithm}
	\DontPrintSemicolon 
	$function\ getAdjectiveImageSchema(Synset\ sense)\ \{$;
	
	$\ \ \ \ breadthFirstSearch(sense, allWordnetSenses, allSynonymEdges)$;
	
	$\}$
	
	$function\ getAdverbImageSchema(Synset\ sense)\ \{$;
	
	$\ \ \ \ adjectiveSense \gets traverseOneHop(sense, allWordnetSenses, allDerivedFromAdjectiveEdges)$;
	
	$\ \ \ \ return getAdjectiveImageSchema(adjectiveSense)$;
	
	$\}$
	
	$function\ getVerbImageSchema(Synset\ sense)\ \{$;
	
	$\ \ \ \ breadthFirstSearch(sense, allWordnetSenses, allHypernymEdges + allTroponymEdges)$;
	
	$\}$
	
	$function\ getNounImageSchema(Synset\ sense)\ \{$;
	
	$\ \ \ \ breadthFirstSearch(sense, allWordnetSenses, allHypernymEdges)$;
	
	$\}$

	\caption{The proposed computation of image schema}
	\label{algo:proposed_image_schema_algo}
\end{algorithm}

\begin{table}
	\begin{center}
		\captionsetup{justification=centering}
		\caption{WordNet 3.0\textquotesingle s unordered connected sense count}
		\label{tab:wordnet_unordered_connected_sense_count}
		\begin{tabular}{|c|c|} 
			\hline
			Type & Count \\
			\hline
			Between nouns (hypernym) & 75850 \\
			Between verbs (hypernym \& troponym) & 13238 \\
			Between adjectives (synonym) & 10693 \\
			\hline
		\end{tabular}
	\end{center}
\end{table}

\section{Proposed Method} \label{sec:proposed_method}

One thing we can easily see is that there are far more English words than there are image schemas. For example, Ravenet et al~\cite{ravenet2018automatic,ravenet2018automating} list only 25 image schemas. Therefore, many different words are necessarily mapped to the same image schema. Considering that those which are mapped to the same image schema should have similar meanings, then the corresponding vectors should also be close to each other, and thus they should form a cluster. Therefore, to answer our first research question (i.e. how can we use word embedding to represent image schema as vector ?), we will use the centroid of the word embedding vectors which belong to the same image schema as the embedding of the image schema.

There are multiple word embedding techniques which can map a word into a vector (see Section \ref{sec:word-embeddings}). BERT is a popular word embedding for a general purpose natural language processing application, and thus BERT is a reasonable choice. However, considering that Ravenet et al~\cite{ravenet2018automatic,ravenet2018automating} use WordNet to map the words into image schemas, a more suitable word embedding might be one which also takes into account the WordNet similarity. SenseBERT is one such embedding technique. Therefore, we will try both BERT and SenseBERT and we will compare their results.

The image schema tagging in Ravenet et al\textquotesingle s algorithm is done per phrase (see Algorithm \ref{algo:existing_meaning_miner_algo} Line \ref{algo_line:per_phrase_loop}). A sentence is split into several phrases (noun phrase, verb phrase, etc.), and then there is at most one image schema per phrase. However, we know from which word the image schema comes from (see Algorithm \ref{algo:existing_meaning_miner_algo} Line \ref{algo_line:lemma_loop}). Therefore, we try two variations of the way to get the vector: by averaging all words in the phrase, or by getting the vector of the word where the image schema comes from.

There are two possible distance/dissimilarity metrics to measure the dissimilarity between two embedding vectors, namely euclidean distance and cosine dissimilarity. Both metrics are often used to measure dissimilarity between word embedding vectors. We will try both metrics to find out if they yield different results.

Having two different word embedding techniques as the possible choices leads us to our second research question: how do we compare the word embedding we choose against the alternative? To answer this question, we will find out which word embedding method gives a better clustering behavior. For that, we use two measures. The first one is cluster purity and the second one is comparing the intra-cluster against the inter-cluster distances.

\begin{figure}
    \centering
	\includegraphics[scale=1.2]{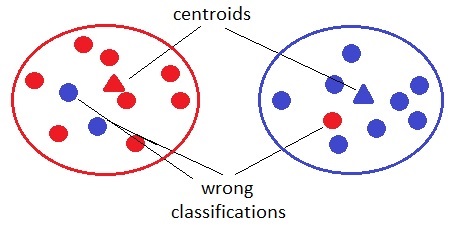}
	\caption{An illustration of the notion of cluster purity. The different colors (i.e. red and blue) represent different image schemas, the small solid circles represent vectors, the triangles represent the centroids, and the large hollow circles represent classifications. The three wrongly classified vectors are misclassified because they are closer to another centroid than to their own centroid. The cluster is purer if there are less misclassified vectors.}
	\label{fig:cluster_purity_example}
\end{figure}

A better clustering behavior yields purer clusters. That means, vectors which belong to a certain image schema should be closer to the centroid of that image schema than to any other centroids. Equivalently, vectors which do not belong to a certain image schema should not have their closest centroid to be the centroid of that image schema. For that, we do a ``classification'' by using the nearest cluster centroid. We ``classify'' each vector according to the nearest centroid. If the nearest centroid is indeed the centroid of its image schema, then we consider it as a correct classification, or else it is a wrong classification. The illustration of this cluster purity notion is furnished in Figure \ref{fig:cluster_purity_example}. According to the results of these classifications, we measure the $F_1$ score of each class. This calculation is done in one-vs-rest manner. That means, when we calculate the $F_1$ score of ``OBJECT'' image schema, we do a binary classification of ``OBJECT'' against all other image schemas. After that, we get the multi-class $F_1$ score from the weighted average of the $F_1$ scores of each class. The weight is proportional to the number of vectors in that image schema/cluster. This weighting is to take into account the fact that some image schemas are more frequent than the others. In this measurement, a higher $F_1$ score is better. The formula is shown in Formula \ref{eq:f1_score}. In that formula, $TP$ stands for the number of ``True Positive''s, $FP$ stands for the number of ``False Positive''s, and $FN$ stands for the number of ``False Negative''s of the aforementioned classification. The $\frac{|v \in IS|}{|v|}$ multiplier in the second line of the formula is the weight of the class\textquotesingle s $F_1$ score.

\begin{figure}
    \centering
	\includegraphics[scale=1.8]{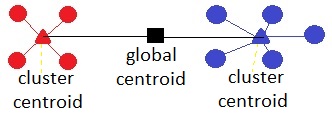}
	\caption{An illustration of the notion of cluster inertia. The different colors (i.e. red and blue) represent different image schemas, the circles represent vectors, the triangles represent the cluster centroids, the black rectangle represents the global centroid, the red or blue lines represent the intra-cluster distances, and the black lines represent the inter-cluster distances. The inertia score is higher if the black lines (i.e. the intra-cluster distances) are longer than the red or blue lines (i.e. the intra-cluster distances)}
	\label{fig:cluster_inertia_example}
\end{figure}

The second measure is to compare the intra-cluster against the inter-cluster distances. Basically, the distances within a cluster should be closer than the distance beyond one cluster. For this, we calculate the inertia score (Formula \ref{eq:inertia_score}). Specifically, we compare the distance between the centroids of each image schema to the global centroid (i.e. the center of all vectors) against the distance between each data point to its cluster centroid. If the distance between each cluster centroid to the global centroid is large and the distance between the individual vectors to their respective centroid is small, then the inertia score will be high. The illustration of this metric is provided in Figure \ref{fig:cluster_inertia_example}. In our measure, a higher inertia score signifies better clustering. We weight each clusters/image schemas linearly to the number of vectors in that image schema to take into account the fact that some image schemas are more frequent than the others.

\begin{equation}
\label{eq:f1_score}
\left.\begin{aligned}
F_{1_{IS}} = \frac{TP}{TP+\frac{FP+FN}{2}} \\
F_1 = \sum_{IS} \frac{|v \in IS|}{|v|} F_{1_{IS}} \\ 
\end{aligned}\right.
\end{equation}

\begin{equation}
\label{eq:inertia_score}
inertia = \sum_{IS} \frac{|v \in IS| \times d(c_{IS}, c_{global})}{\sum_{v \in IS} d(v, c_{IS})}
\end{equation}

Once each image schema can be represented as a vector, then the notion of distance becomes sensible. Previously, we have a notion of different image schemas, but we cannot tell if some image schemas are closer or more similar to each other than to the others. However, if each image schema is represented by a vector, then we can calculate the distance between them. This is to answer our third research question: which image schemas are close to each other? We will measure their distances and also do a hierarchical clustering to show how similar image schemas can be merged. We use two measures to calculate the distance between different image schemas.

In the first measure (Formula \ref{eq:inter_centroid_distance}), we calculate the distance between their respective centroids. If the two centroids are near, then the two image schemas are considered to be similar.

In the second measure, we measure the confusion between the individual vectors of each image schema (Formula \ref{eq:inter_confusion_distance}). Specifically, given two image schemas $IS_1$ and $IS_2$, we measure how many individual vectors belonging to image schema $IS_1$ which are closer to the centroid of $IS_2$ and vice versa. The more numerous such vectors are (i.e. the more ``confused'' the vectors are), the closer the two image schemas are. This is essentially the inverse of the cluster purity measure. We normalize the number of the ``confused'' vectors against the size of each image schema. Therefore, this confusion dissimilarity metrics works at the level of proportion. For example, if only 30\% of $IS_1$\textquotesingle s vectors are closer to the centroid of $IS_1$ (than to the centroid of $IS_2$) and only 40\% of $IS_2$\textquotesingle s vectors are closer to the centroid of $IS_2$ (than to the centroid of $IS_1$), then the confusion is $(0.3 + 0.4) / 2 = 0.35$. The lower the score, the closer the two image schemas are.

\begin{figure}
    \centering
	\includegraphics[scale=1]{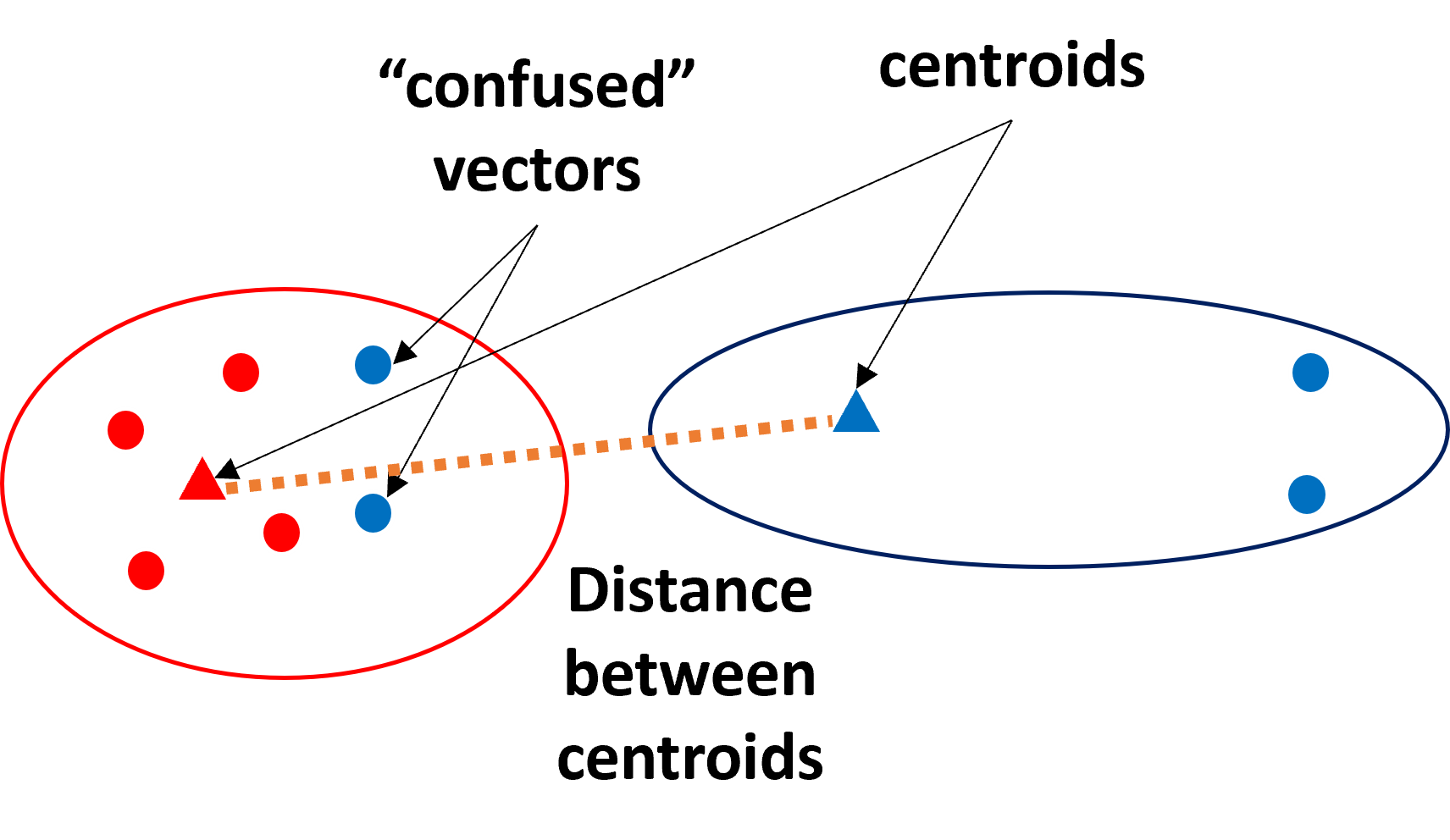}
	\caption{An illustration of image schema distance/dissimilarity. The different colors (i.e. red and blue) represent different image schemas, the circles represent vectors, the triangles represent the cluster centroids, and the orange dotted line represents the distance between the two centroids. Two blue vectors are nearer to the red centroid, and thus they are ``confused''. On the distance/dissimilarity metric which uses the inter-centroid distance (Formula \ref{eq:inter_centroid_distance}), the two image schemas are closer to each other if the dotted orange line is shorter. On the distance/dissimilarity metric which uses the confusion (Formula \ref{eq:inter_confusion_distance}), the two image schemas are closer to each other if there are more ``confused'' vectors.
	}
	\label{fig:image_schema_distance_example}
\end{figure}

Both of those dissimilarity metrics are illustrated in Figure \ref{fig:image_schema_distance_example}. The confusion-based metrics is essentially the inverse of the cluster purity measure. The main difference between those two metrics is that the confusion dissimilarity metrics takes into account the dispersion of the vectors. For example, in Figure \ref{fig:image_schema_distance_example}, the red cluster is more compact than the blue cluster. Two of the blue vectors are actually closer to the red centroid than to the blue centroid.

\begin{equation}
\label{eq:inter_centroid_distance}
d_{centroid}(IS_1, IS_2) = d(c_{IS_1}, c_{IS_2})
\end{equation}

\begin{equation}
\label{eq:inter_confusion_distance}
d_{confusion}(IS_1, IS_2) = \frac{1}{2} (\frac{|v \mid v \in IS_1 \land d(v, c_{IS_1} > d(v, c_{IS_2}))|}{|v \in IS_1|} + \frac{|v \mid v \in IS_2 \land d(v, c_{IS_1} < d(v, c_{IS_2}))|}{|v \in IS_2|})
\end{equation}

\section{Experiment} \label{sec:experiment}

In this section, we compare the effect of the various choices of ``parameters'' for the calculation of the image schema embedding vectors. Those ``parameter'' choices are the choice of word embedding technique (either BERT or SenseBERT) and the embedding source (either we consider only the embedding vector from the word where the image schema comes from or we average the embedding vectors from the whole phrase). With the image schema embedding vectors known, we measure the relative pairwise dissimilarities between the different image schemas. We also show some visualizations to illustrate the relative distances between the different image schemas.

We use the Stanford Natural Language Inference (SNLI)~\cite{bowman2015large} corpus as the source of our sentences. This corpus has 652,505 sentences in English. First, we run the modified Ravenet et al\textquotesingle s algorithm (whose modification is explained in Section \ref{sec:limitation_of_ravenet_algo}) to extract the image schema from those sentences and we mark the word where the image schema comes from. For the purpose of illustration, we furnish a few examples of sentences and the image schema in Table \ref{tab:sentence-image_schema-examples}. We furnish the statistics of the extracted image schema from the corpus in Table \ref{tab:image_schema_count_and_their_proportion}. After we have extracted the image schemas, we run both BERT and SenseBERT to get the embedding of all the words in the corpus.

\begin{table}
	\begin{center}
		\captionsetup{justification=centering}
		\caption{Examples of sentence and image schema} \label{tab:sentence-image_schema-examples}
		\begin{tabular}{|c|c|}
			\hline
			Sentence & Image Schema \\
			\hline
            The men are fighting outside a deli. & ``deli'' is WHOLE \\
            Two kids in numbered jerseys wash their hands. & ``hands'' is OBJECT \\
            The woman\textquotesingle s hands are empty & ``empty'' is EMPTY \\
            A woman is talking on the phone while standing next to a dog. & ``next'' is RIGHT \\
			\hline
		\end{tabular}
	\end{center}
\end{table}

\begin{table}
	\begin{center}
		\captionsetup{justification=centering}
		\caption{Image schema count and their proportion} \label{tab:image_schema_count_and_their_proportion}
		\begin{tabular}{|c|c|c|}
			\hline
			Image Schema & Count & Proportion \\
			\hline
			ATTRACTION & 1368 & 0.0997\% \\
			BACK & 2104 & 0.153\% \\
			BIG & 13863 & 1.010\% \\
			CONTAINER & 11147 & 0.812\% \\
			CONTRAST & 234 & 0.017\% \\
			DOWN & 3284 & 0.239\% \\
			EMPTY & 1659 & 0.121\% \\
			ENABLEMENT & 7980 & 0.581\% \\
			FAR & 8830 & 0.643\% \\
			FRONT & 1459 & 0.106\% \\
			FULL & 1011 & 0.074\% \\
			IN & 143151 & 10.431\% \\
			INTERVAL & 69848 & 5.090\% \\
			ITERATION & 406 & 0.030\% \\
			LEFT & 1534 & 0.112\% \\
			LINK & 19387 & 1.413\% \\
			MERGING & 6146 & 0.448\% \\
			NEAR & 10203 & 0.743\% \\
			OBJECT & 550542 & 40.116\% \\
			OUT & 13594 & 0.991\% \\
			RIGHT & 8469 & 0.617\% \\
			SMALL & 3087 & 0.225\% \\
			SURFACE & 5932 & 0.432\% \\
			UP & 25226 & 1.838\% \\
			WHOLE & 461918 & 33.658\% \\
			\hline
		\end{tabular}
	\end{center}
\end{table}

In our first experiment, we measure the clusters\textquotesingle $F_1$ score/purity (see Formula \ref{eq:f1_score}) and inertia (see Formula \ref{eq:inertia_score}) to compare BERT and SenseBERT on the resulting clustering behavior. We permute through the options of cosine dissimilarity - euclidean distance and embedding comes from the word - embedding is averaged through the phrase. 

We furnish in Table \ref{tab:intertia_by_averaging} the inertia when we calculate the value of each vector as the average BERT/SenseBERT embedding vector of all the words in the phrase. On the other hand, in Table \ref{tab:intertia_per_word}, we show the corresponding values if each vector is the BERT/SenseBERT embedding vector of the word where the image schema comes from.

\begin{table}
	\begin{center}
		\captionsetup{justification=centering}
		\caption{Inertia measure with each vector is calculated by averaging from all words in the phrase} \label{tab:intertia_by_averaging}
		\begin{tabular}{|c|c|c|}
			\hline
			& By Euclidean Distance & By Cosine Dissimilarity \\
			\hline
			BERT & 0.047 & 0.072 \\
			SenseBERT & 0.052 & 0.104 \\
			\hline
		\end{tabular}
	\end{center}
\end{table}

\begin{table}
	\begin{center}
		\captionsetup{justification=centering}
		\caption{Inertia measure with each vector comes from the word where the image schema comes from} \label{tab:intertia_per_word}
		\begin{tabular}{|c|c|c|}
			\hline
			& By Euclidean Distance & By Cosine Dissimilarity \\
			\hline
			BERT & 0.048 & 0.082 \\
			SenseBERT & 0.078 & 0.158 \\
			\hline
		\end{tabular}
	\end{center}
\end{table}

Similarly, we furnish in Tables \ref{tab:f1_by_averaging} and \ref{tab:f1_per_word} the $F_1$ score when we calculate the value of each vector as the average BERT/SenseBERT embedding vector of all the words in the phrase and when the value if each vector is the BERT/SenseBERT embedding vector of the word where the image schema comes from.

\begin{table}
	\begin{center}
		\captionsetup{justification=centering}
		\caption{$F_1$ score with each vector is calculated by averaging from all words in the phrase} \label{tab:f1_by_averaging}
		\begin{tabular}{|c|c|c|}
			\hline
			& By Euclidean Distance & By Cosine Dissimilarity \\
			\hline
			BERT & 0.439 & 0.526 \\
			SenseBERT & 0.613 & 0.609 \\
			\hline
		\end{tabular}
	\end{center}
\end{table}

\begin{table}
	\begin{center}
		\captionsetup{justification=centering}
		\caption{$F_1$ score with each vector comes from the word where the image schema comes from} \label{tab:f1_per_word}
		\begin{tabular}{|c|c|c|}
			\hline
			& By Euclidean Distance & By Cosine Dissimilarity \\
			\hline
			BERT & 0.627 & 0.619 \\
			SenseBERT & 0.705 & 0.697 \\ 
			\hline
		\end{tabular}
	\end{center}
\end{table}

\begin{figure}
    \centering
	\includegraphics[scale=0.7]{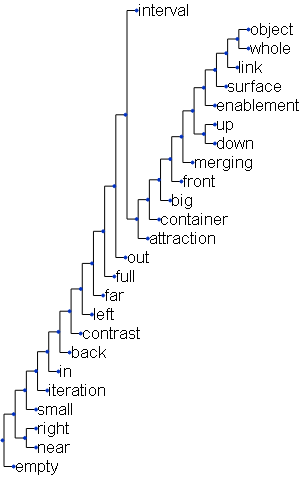}
	\caption{The hierarchical clustering of image schemas in BERT and euclidean distance (between two vectors) with the centroid distance (between two image schemas). Nearby image schemas are merged earlier while outlying ones are merged later.}
	\label{fig:tree_euclidean_word_bert_centroid}
\end{figure}

\begin{figure}
    \centering
	\includegraphics[scale=0.7]{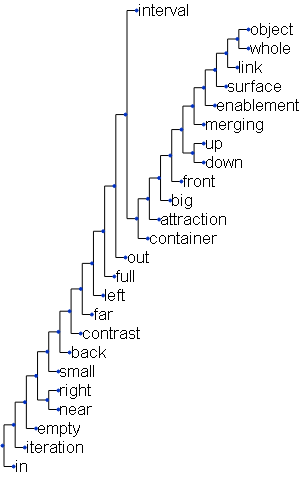}
	\caption{The hierarchical clustering of image schemas in BERT and cosine dissimilarity (between two vectors) with the centroid distance (between two image schemas). Nearby image schemas are merged earlier while outlying ones are merged later.}
	\label{fig:tree_cosine_word_bert_centroid}
\end{figure}

\begin{figure}
    \centering
	\includegraphics[scale=0.7]{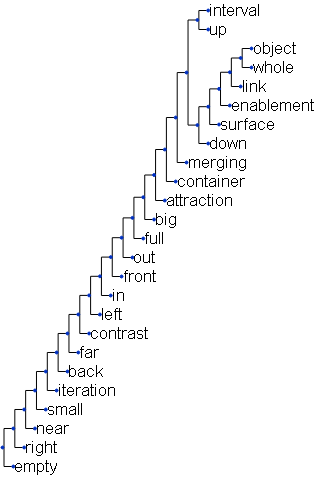}
	\caption{The hierarchical clustering of image schemas in SenseBERT and euclidean distance (between two vectors) with the centroid distance (between two image schemas). Nearby image schemas are merged earlier while outlying ones are merged later.}
	\label{fig:tree_euclidean_word_sensebert_centroid}
\end{figure}

\begin{figure}
    \centering
	\includegraphics[scale=0.7]{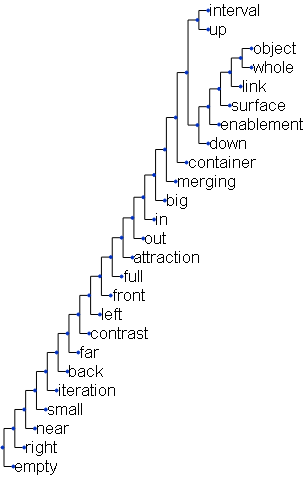}
	\caption{The hierarchical clustering of image schemas in SenseBERT and cosine dissimilarity (between two vectors) with the centroid distance (between two image schemas). Nearby image schemas are merged earlier while outlying ones are merged later.}
	\label{fig:tree_cosine_word_sensebert_centroid}
\end{figure}

\begin{figure}
    \centering
	\includegraphics[scale=0.7]{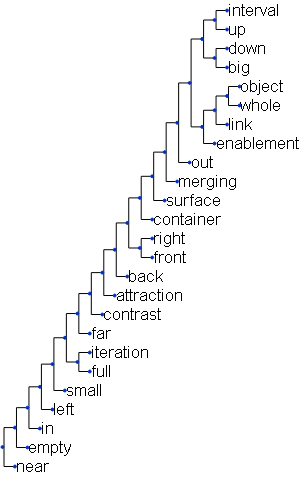}
	\caption{The hierarchical clustering of image schemas in BERT and euclidean distance (between two vectors) with the confusion dissimilarity (between two image schemas). Nearby image schemas are merged earlier while outlying ones are merged later.}
	\label{fig:tree_euclidean_word_bert_confusion}
\end{figure}

\begin{figure}
    \centering
	\includegraphics[scale=0.7]{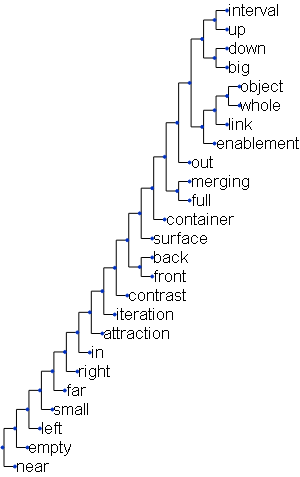}
	\caption{The hierarchical clustering of image schemas in BERT and cosine distance (between two vectors) with the confusion distance (between two image schemas). Nearby image schemas are merged earlier while outlying ones are merged later.}
	\label{fig:tree_cosine_word_bert_confusion}
\end{figure}

\begin{figure}
    \centering
	\includegraphics[scale=0.7]{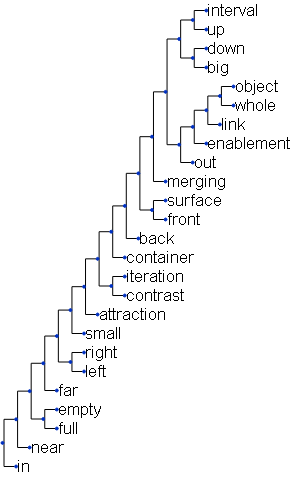}
	\caption{The hierarchical clustering of image schemas in SenseBERT and euclidean distance (between two vectors) with the confusion dissimilarity (between two image schemas). Nearby image schemas are merged earlier while outlying ones are merged later.}
	\label{fig:tree_euclidean_word_sensebert_confusion}
\end{figure}

\begin{figure}
    \centering
	\includegraphics[scale=0.7]{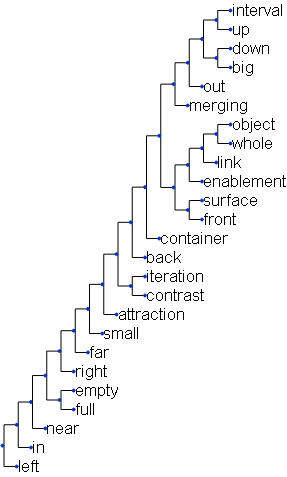}
	\caption{The hierarchical clustering of image schemas in SenseBERT and cosine dissimilarity (between two vectors) with the confusion dissimilarity (between two image schemas). Nearby image schemas are merged earlier while outlying ones are merged later.}
	\label{fig:tree_cosine_word_sensebert_confusion}
\end{figure}

In our second experiment, we want to find out which image schemas are close to each other. However, we have already represented the image schemas as clusters. Therefore, we measure the distances between their respective clusters as a proxy of similarities/differences between the different image schemas. For this, we measure the inter-centroid distances (see Formula \ref{eq:inter_centroid_distance}) and the confusion rates (see Formula \ref{eq:inter_confusion_distance}). It is useful to remember that the main difference between the inter-centroid distance metrics and the confusion metrics is that the confusion metrics takes into account the dispersion of the cluster while the inter-centroid distance metrics does not. We permute through the options of BERT - SenseBERT and cosine dissimilarity - euclidean distance. We furnish the five closest pairs of image schemas according to those parameters in Tables \ref{tab:closest_image_schemas_bert_centroid}, \ref{tab:closest_image_schemas_sensebert_centroid}, \ref{tab:closest_image_schemas_bert_confusion}, and \ref{tab:closest_image_schemas_sensebert_confusion}. We also show the hierarchical clusters of the image schemas according to the same permutation of parameters. In this hierarchical clustering, we merge the closest pair of image schemas at each step until there is only one remaining. In effect, nearby image schemas will be merged early while ``remote'' image schemas will be merged the last. The hierarchical clusters are displayed in Figures \ref{fig:tree_euclidean_word_bert_centroid}, \ref{fig:tree_cosine_word_bert_centroid}, \ref{fig:tree_euclidean_word_sensebert_centroid}, \ref{fig:tree_cosine_word_sensebert_centroid}, \ref{fig:tree_euclidean_word_bert_confusion}, \ref{fig:tree_cosine_word_bert_confusion}, \ref{fig:tree_euclidean_word_sensebert_confusion}, and \ref{fig:tree_cosine_word_sensebert_confusion}

\begin{table}
	\begin{center}
		\captionsetup{justification=centering}
		\caption{Five closest pairs of image schemas in BERT with the centroid distance} \label{tab:closest_image_schemas_bert_centroid}
		\begin{tabular}{|c|c|c|}
			\hline
			Closeness & BERT / Euclidean & BERT / Cosine \\
			Rank & Distance & Dissimilarity \\
			\hline
			1 & OBJECT-WHOLE & OBJECT-WHOLE \\
			2 & LINK-OBJECT & LINK-OBJECT \\ 
			3 & LINK-WHOLE & LINK-WHOLE \\ 
			4 & ENABLEMENT-OBJECT & ENABLEMENT-OBJECT \\ 
			5 & OBJECT-SURFACE & OBJECT-SURFACE \\ 
			\hline
		\end{tabular}
	\end{center}
\end{table}

\begin{table}
	\begin{center}
		\captionsetup{justification=centering}
		\caption{Five closest pairs of image schemas in SenseBERT with the centroid distance} \label{tab:closest_image_schemas_sensebert_centroid}
		\begin{tabular}{|c|c|c|}
			\hline
			Closeness & SenseBERT / Euclidean & SenseBERT / Cosine \\
			Rank & Distance & Distance \\
			\hline
			1 & OBJECT-WHOLE & OBJECT-WHOLE \\
			2 & LINK-OBJECT & LINK-OBJECT \\ 
			3 & ENABLEMENT-OBJECT & LINK-WHOLE \\ 
			4 & LINK-WHOLE & ENABLEMENT-OBJECT \\ 
			5 & OBJECT-SURFACE & OBJECT-SURFACE \\ 
			\hline
		\end{tabular}
	\end{center}
\end{table}

\begin{table}
	\begin{center}
		\captionsetup{justification=centering}
		\caption{Five closest pairs of image schemas in BERT with the confusion distance} \label{tab:closest_image_schemas_bert_confusion}
		\begin{tabular}{|c|c|c|}
			\hline
			Closeness & BERT / Euclidean & BERT / Cosine \\
			Rank & Distance & Dissimilarity \\
			\hline
			1 & OBJECT-WHOLE & OBJECT-WHOLE \\
			2 & INTERVAL-UP & LINK-OBJECT \\
			3 & LINK-OBJECT & INTERVAL-UP \\
			4 & LINK-WHOLE & LINK-WHOLE \\
			5 & ENABLEMENT-OBJECT & BIG-UP \\
			\hline
		\end{tabular}
	\end{center}
\end{table}

\begin{table}
	\begin{center}
		\captionsetup{justification=centering}
		\caption{Five closest pairs of image schemas in SenseBERT with the confusion dissimilarity} \label{tab:closest_image_schemas_sensebert_confusion}
		\begin{tabular}{|c|c|c|}
			\hline
			Closeness & SenseBERT / Euclidean & SenseBERT / Cosine \\
			Rank & Distance & Dissimilarity \\
			\hline
			1 & OBJECT-WHOLE & OBJECT-WHOLE \\
			2 & INTERVAL-UP & INTERVAL-UP \\ 
			3 & BIG-DOWN & BIG-DOWN \\ 
			4 & LINK-OBJECT & LINK-OBJECT \\ 
			5 & BIG-UP & BIG-UP \\ 
			\hline
		\end{tabular}
	\end{center}
\end{table}

\begin{figure}
    \centering
	\includegraphics[scale=0.1]{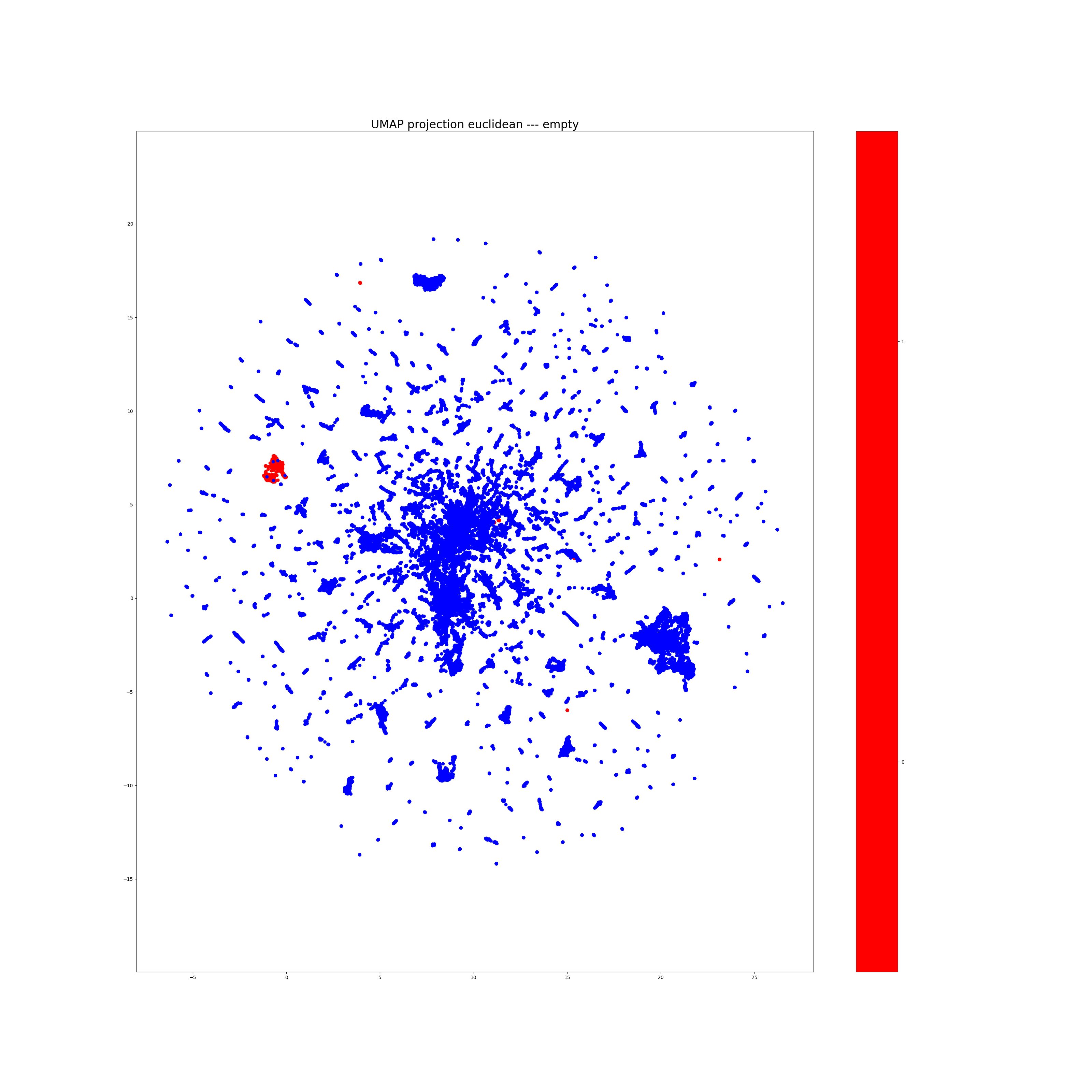}
	\caption{The UMAP visualization of all vectors in the samples with BERT and euclidean distance (between two vectors). The red points belong to EMPTY image schema and the blue points belong to any other image schema}
	\label{fig:visualization_bert_euclidean_empty}
\end{figure}

\begin{figure}
    \centering
	\includegraphics[scale=0.1]{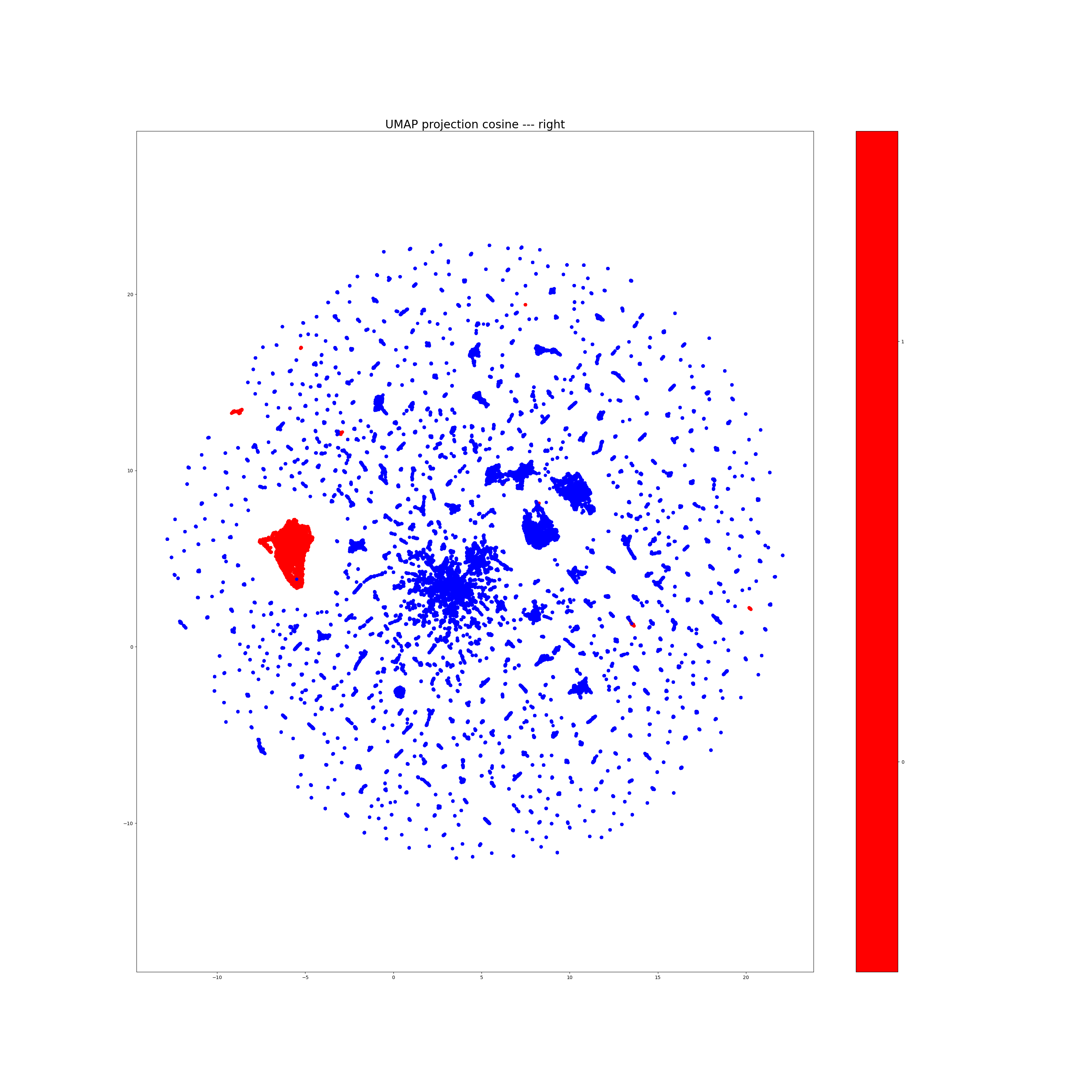}
	\caption{The UMAP visualization of all vectors in the samples with SenseBERT and cosine dissimilarity (between two vectors). The red points belong to RIGHT image schema and the blue points belong to any other image schema}
	\label{fig:visualization_sensebert_cosine_right}
\end{figure}

\begin{figure}
    \centering
	\includegraphics[scale=0.1]{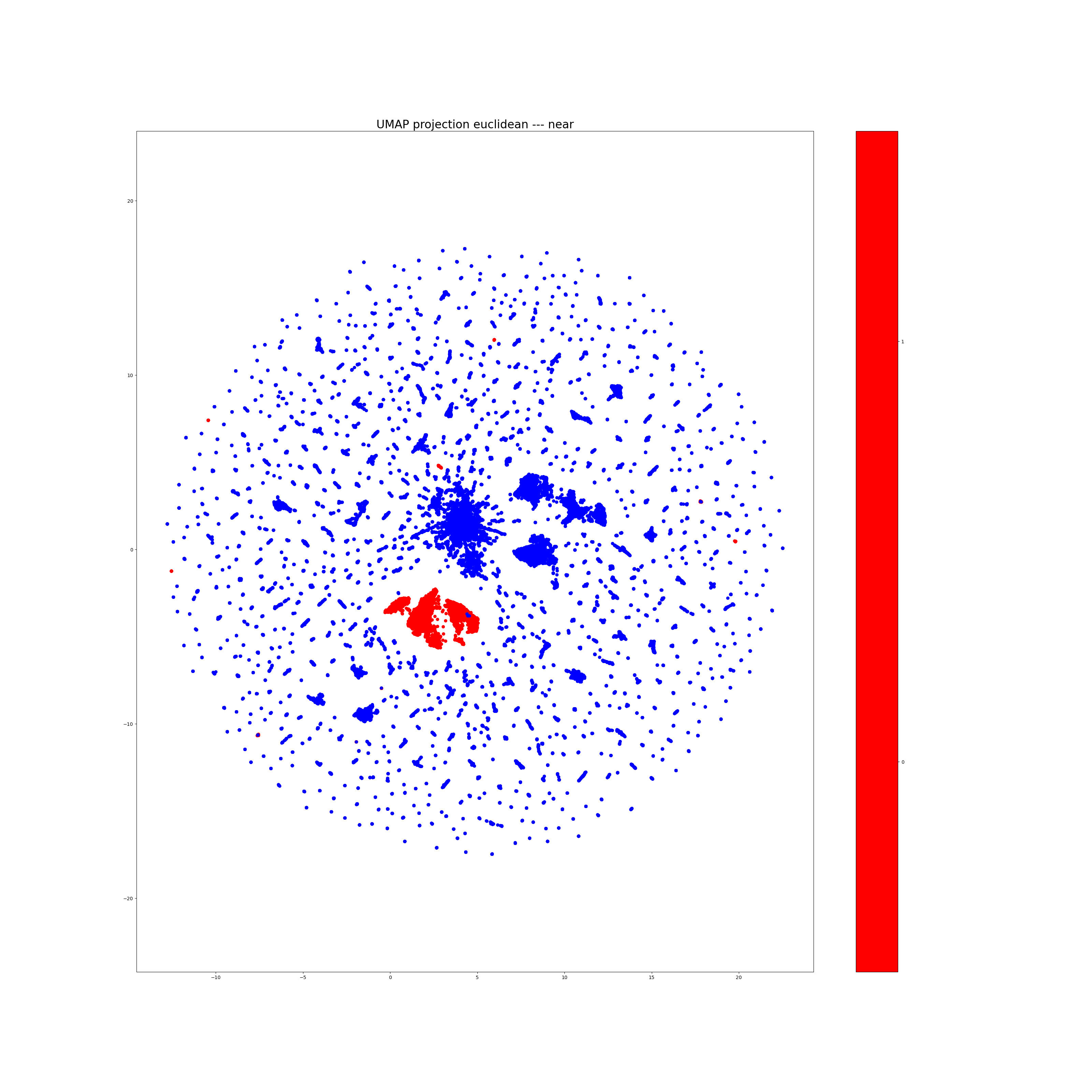}
	\caption{The UMAP visualization of all vectors in the samples with SenseBERT and euclidean distance (between two vectors). The red points belong to NEAR image schema and the blue points belong to any other image schema}
	\label{fig:visualization_sensebert_euclidean_near}
\end{figure}

We also display the UMAP visualizations~\cite{mcinnes2018umap} of three cases, namely EMPTY image schema in BERT and euclidean distance (Figure \ref{fig:visualization_bert_euclidean_empty}), RIGHT image schema in SenseBERT and cosine dissimilarity (Figure \ref{fig:visualization_sensebert_cosine_right}), and NEAR image schema in SenseBERT and euclidean distance (Figure \ref{fig:visualization_sensebert_euclidean_near}). We choose them because according to the hierarchical clusters when we use the centroid distance, they are among the last to be merged, which suggest that they are a relative outlier among all the vectors. It should be noted that due to the large number of the vectors, we cannot run the UMAP algorithm with our entire data. Therefore, we sample the data to make the computation tractable.

\section{Discussion} \label{sec:discussion}

We can see from Tables \ref{tab:intertia_by_averaging}, \ref{tab:intertia_per_word}, \ref{tab:f1_by_averaging}, and \ref{tab:f1_per_word}, SenseBERT embeddings show a better clustering behavior. It holds true whether we measure the clustering by using inertia or by using $F_1$ score. It also holds true whether we measure the distance/dissimilarity by using euclidean distance or by using cosine dissimilarity. It also holds true whether the vector values are obtained by averaging the BERT/SenseBERT embedding vectors of all the words in the phrase or by taking the BERT/SenseBERT embedding vector of the word which is tagged with the image schema. This is likely caused by the fact that SenseBERT is explicitly trained by using WordNet, which the image schema computation also uses. Therefore, if we want to represent an image schema as a vector, SenseBERT is more suitable as the base than BERT is. We can use the centroids of the clusters as the embedding vector for each image schema.

We also observe in the same four tables that the inertias and the $F_1$ scores are higher when we take the vector values from only the word which is tagged with the image schema instead of when we average all the words in the phrase. This is likely caused by the fact that the rests of the phrase are quite similar across various image schemas, which cause the differences between different image schemas (i.e. different clusters) become less pronounced than when we take only the words which are tagged with the image schemas.

We can see the lists of the closest pairs of image schemas in Tables \ref{tab:closest_image_schemas_bert_centroid}, \ref{tab:closest_image_schemas_sensebert_centroid}, \ref{tab:closest_image_schemas_bert_confusion}, and \ref{tab:closest_image_schemas_sensebert_confusion}. Interestingly, OBJECT-WHOLE is the closest pair in all cases. Besides that, we can see in Table \ref{tab:image_schema_count_and_their_proportion} that OBJECT and WHOLE together form a large majority of our vectors: more than 70\%. It suggests that both image schemas are too general and should be split and refined further.

We can also see in Tables \ref{tab:closest_image_schemas_bert_centroid} and \ref{tab:closest_image_schemas_sensebert_centroid} that both BERT and SenseBERT give the same top-five closest pairs (albeit in a slightly different order) when we use the centroid distance to measure the distance/dissimilarity between two image schemas. However, either OBJECT or WHOLE (or both) appear in all those pairs of image schemas. We also know that the centroids of OBJECT and WHOLE are close to each other. We also see in the dendrograms of the hierarchical clusters when we use the centroid distance to measure the distance/dissimilarity between two image schemas (Figures \ref{fig:tree_euclidean_word_bert_centroid}, \ref{fig:tree_cosine_word_bert_centroid}, \ref{fig:tree_euclidean_word_sensebert_centroid}, and \ref{fig:tree_cosine_word_sensebert_centroid}) that OBJECT and WHOLE are merged first, then other image schema is merged there one by one. It suggests that both OBJECT and WHOLE are located relatively in the center and other image schemas ``radiate'' from there.

On the other hand, we see in Tables \ref{tab:closest_image_schemas_bert_confusion} and \ref{tab:closest_image_schemas_sensebert_confusion} where we use the confusion dissimilarity metrics that the top-five closest pairs are more diverse: not all of them include either OBJECT or WHOLE. We find INTERVAL-UP, BIG-UP, and BIG-DOWN. The difference with the aforementioned case when we use the centroid distance might be caused by the dispersion of the vectors away from their respective centroid. The centroids of INTERVAL and UP are likely to be not so near, but there are many of INTERVAL and UP vectors which are near to the centroid of the other image schema. These results also suggest that, those pairs of image schema are likely often ``confused with each other'' and probably they should be split and refined further.

We also see in the dendrograms of the hierarchical clusters when we use the confusion to measure the distance/dissimilarity between two image schemas (Figures \ref{fig:tree_euclidean_word_bert_confusion}, \ref{fig:tree_cosine_word_bert_confusion}, \ref{fig:tree_euclidean_word_sensebert_confusion}, and \ref{fig:tree_cosine_word_sensebert_confusion}) that there are many cases of image schemas other than OBJECT and WHOLE merged quite early: INTERVAL-UP (Figures \ref{fig:tree_euclidean_word_bert_confusion}, \ref{fig:tree_cosine_word_bert_confusion}, \ref{fig:tree_euclidean_word_sensebert_confusion}, and \ref{fig:tree_cosine_word_sensebert_confusion}), BIG-DOWN (Figures \ref{fig:tree_euclidean_word_bert_confusion}, \ref{fig:tree_cosine_word_bert_confusion}, \ref{fig:tree_euclidean_word_sensebert_confusion}, and \ref{fig:tree_cosine_word_sensebert_confusion}), MERGING-FULL (Figure \ref{fig:tree_cosine_word_bert_confusion}), and SURFACE-FRONT (Figures \ref{fig:tree_euclidean_word_sensebert_confusion} and \ref{fig:tree_cosine_word_sensebert_confusion}). These results are more diverse than when we use the centroid distance where the hierarchical clusters show that the OBJECT and WHOLE are to be merged together and then other image schemas are to be merged into the OBJECT-WHOLE cluster one by one. The more diverse results when we use the confusion dissimilarity metrics is likely because the confusion dissimilarity metrics takes into account the vectors which are far from their respective centroid and there are many such vectors.

We try both euclidean distance and cosine dissimilarity metrics in our various experiments. From strictly mathematical point of view, the euclidean and cosine metrics are not interchangeable. For example, let $A$, $B$, and $C$ be points in a 2-dimension vector space where $A = (1, 0)$, $B = (10, 10)$, and $C = (1, 2)$. In this example, $dissimilarity_{cosine}(A, B) < dissimilarity_{cosine}(A, C)$ but $distance_{euclidean}(A, B) > distance_{euclidean}(A, C)$. Yet, in the context of word embedding works, both metrics are regularly used and are often treated as if they were interchangeable. We are also not aware of any formal proof to support their interchangeability in the context of word embedding. Indeed, this is the reason we try both the euclidean and cosine metrics. In the experiments where we evaluate the clustering behaviors (see Tables \ref{tab:intertia_by_averaging}, \ref{tab:intertia_per_word}, \ref{tab:f1_by_averaging}, and \ref{tab:f1_per_word}), we see that whether we use euclidean or cosine metrics, SenseBERT is always better than BERT. In the experiment where we measure the closest pairs of image schema when we use BERT and centroid distance between two image schemas (see Table \ref{tab:closest_image_schemas_bert_centroid}), we see that the five closest image schema pairs are exactly the same. When we use SenseBERT and centroid distance (see Table \ref{tab:closest_image_schemas_sensebert_centroid}), we see that the five closest image schema pairs are the same although ENABLEMENT-OBJECT and LINK-WHOLE swap their ranks (the 3rd and the 4th respectively). When we use BERT and confusion dissimilarity metrics (see Table \ref{tab:closest_image_schemas_bert_confusion}), we see that the four closest image schema pairs are the same except that INTERVAL-UP and LINK-OBJECT swap their ranks (the 2nd and the 3rd respectively). The fifth closest pair (i.e. ENABLEMENT-OBJECT and BIG-UP) are different when we use euclidean and cosine metrics. When we use SenseBERT and confusion dissimilarity (see Table \ref{tab:closest_image_schemas_sensebert_confusion}), we see that the five closest image schema pairs are exactly the same. In all the aforementioned experiments where we measure distances/dissimilarities between different image schema pairs, we see that the five closest image schema pairs only differ a bit when we use euclidean or cosine metrics. However, we notice different merging sequences when we use euclidean and cosine metrics, especially after the first few image schemas are merged (see Figures \ref{fig:tree_euclidean_word_bert_centroid}-\ref{fig:tree_cosine_word_bert_centroid}, \ref{fig:tree_euclidean_word_sensebert_centroid}-\ref{fig:tree_cosine_word_sensebert_centroid}, \ref{fig:tree_euclidean_word_bert_confusion}-\ref{fig:tree_cosine_word_bert_confusion}, and \ref{fig:tree_euclidean_word_sensebert_confusion}-\ref{fig:tree_cosine_word_sensebert_confusion}). From all those experiments, we see that the both euclidean and cosine distance/dissimilarity metrics yield similar but not exactly the same results.

An interesting observation from the results of the experiments where we measure the distances/dissimilarities of different image schema pairs is that OBJECT-WHOLE is always the closest pair, whether we use BERT or SenseBERT, centroid or confusion dissimilarity between two image schemas, or euclidean or cosine dissimilarity between two vectors (see Tables \ref{tab:closest_image_schemas_bert_centroid}, \ref{tab:closest_image_schemas_sensebert_centroid}, \ref{tab:closest_image_schemas_bert_confusion}, and \ref{tab:closest_image_schemas_sensebert_confusion}). Besides that, these two image schemas together constitute more than 70\% of all the data points (see Table \ref{tab:image_schema_count_and_their_proportion}). These results suggest that OBJECT and WHOLE image schemas probably should be split and refined further.

Finally, we show three UMAP visualizations: the EMPTY image schema when we use BERT and euclidean distance (Figure \ref{fig:visualization_bert_euclidean_empty}), the RIGHT image schema when we use SenseBERT and cosine dissimilarity (Figure \ref{fig:visualization_sensebert_cosine_right}) and the NEAR image schema when we use SenseBERT and euclidean distance (Figure \ref{fig:visualization_sensebert_euclidean_near}). In all the three cases, those image schemas are among the last to be merged in their respective hierarchical clustering when we use the centroid distance (Figure \ref{fig:tree_euclidean_word_bert_centroid} for EMPTY, Figure \ref{fig:tree_cosine_word_sensebert_centroid} for RIGHT, and Figure \ref{fig:tree_euclidean_word_sensebert_centroid} for NEAR). The fact that they are among the last to be merged suggests that they are a relative outlier, and indeed we see in the three UMAP visualizations that they are away from the dense center.

\section{Conclusion} \label{sec:conclusion}

In this work, we propose a technique to represent image schemas as vectors. As far as we are aware of, this is the first work which addresses that problem. With the image schema representable as vectors, it becomes possible in the future to use image schema on neural network. This is analoguous with word embedding technique which is used to represent words as vectors which in turn allows the use of neural network to solve Natural Language Processing problems. Another effect of being able to represent the image schemas as vectors is that it also becomes possible to calculate the distances between different image schemas. This distance between the two vectors is a proxy of the dissimilarity between the corresponding image schemas. With that, we find out which image schemas are relatively close to each other. We also show visualizations of the relative distances between the different image schemas in the form of hierarchical clusters. In the hierarchical clusters, nearby image schemas are merged early while relatively outlying image schemas are merged the last. Finally, we show visualizations of the clustering behavior of a few image schemas in the form of UMAP visualizations of the data.

\section{Future Work} \label{sec:future_work}

This work is undertaken within the larger project of gesture generation by using machine learning. This work itself is inspired by the work of Ravenet et al~\cite{ravenet2018automatic,ravenet2018automating} which generates metaphoric gestures by using image schema. This work is a representation learning work on image schema. However, in its current state, our representation learning process completely ignores the gestures. For example, we find in our various experiments that OBJECT and WHOLE image schemas are close to each other. Intuitively, that probably means that the gestures corresponding to OBJECT and WHOLE image schemas should also be close to each other. Yet, we do not know if it is indeed the case. Such work obviously requires a gesture representation scheme and a way to measure gesture similarity/difference. However, if we have a gesture representation scheme and gesture data, it might be interesting to integrate the gesture data into the learning process.

In a different direction, it is also an interesting possibility to measure the relative distances of different image schemas in a subjective study and to compare the result against ours. We can probably think of an experiment similar to the one done by Cienki~\cite{cienki2013image} such that we ask the participants to label various messages with their appropriate image schema. If similar number of participants label OBJECT and WHOLE for each messages, then it means OBJECT and WHOLE are similar.

On the minutiae of the technique, as we mention in the Section \ref{sec:limitation_of_ravenet_algo}, our choice of using the Wordnet 1st Sense as the word sense disambiguation techniques is based on the results of Raganato et al\textquotesingle s experiments~\cite{raganato2017word}. However, there have been new word sense disambiguation techniques which were not tested in the Raganato et al\textquotesingle s experiments. It is interesting to try more ``modern'' word sense disambiguation techniques and to find out their effects.

\section*{Acknowledgement}
This project has received funding from the European Union\textquotesingle s Horizon 2020 research and innovation programme under grant agreement No. 769553.  This result dissemination reflects only the authors\textquotesingle s views. The European Commission is not responsible for any use that may be made of the information it contains. We thank Brian Ravenet for his help on many occasions. His help was instrumental for us to understand the minutiae of his algorithm. We also thank Alexandre Pauchet for his helpful feedbacks.

\clearpage

\bibliographystyle{unsrt}  
\bibliography{main}

\end{document}